\documentclass[prl,aps,amssymb,twocolumn,showpacs,floatfix]{revtex4}

\usepackage{graphicx}
\usepackage{dcolumn}
\usepackage{bm}

\begin{document}

\preprint{}

\title{Bright-Exciton Fine Structure and Anisotropic Exchange in CdSe Nanocrystal Quantum Dots}

\author{M.~Furis$^{1}$, H. Htoon$^{2}$, M. A.~Petruska$^{2}$, V.~I.~Klimov$^{2}$,
T.~Barrick$^{1}$, and S.~A.~Crooker$^{1}$}\email{crooker@lanl.gov}

\affiliation{$^{1}$National High Magnetic Field Laboratory, Los Alamos, NM 87545}

\affiliation{$^{2}$Chemistry Division, Los Alamos National Laboratory, Los Alamos, NM 87545}

\date{\today}

\begin{abstract}
We report on polarization-resolved resonant photoluminescence (PL) spectroscopy of bright (spin
$\pm$1) and dark (spin $\pm$2) excitons in colloidal CdSe nanocrystal quantum dots. Using high
magnetic fields to 33 T, we resonantly excite (and selectively analyze PL from) spin-up or
spin-down excitons. At low temperatures ($<$4K) and above $\sim$10 T, the spectra develop a narrow,
circularly polarized peak due to spin-flipped bright excitons. Its evolution with magnetic field
directly reveals a large (1-2 meV), intrinsic fine structure splitting of bright excitons, due to
anisotropic exchange.  These findings are supported by time-resolved PL studies and
polarization-resolved PL from single nanocrystals.
\end{abstract}
\pacs{78.67.Bf, 73.21.La, 73.22.-f, 81.07.Ta}

\maketitle
Semiconductor quantum dots are often regarded as model candidates for ``qubits" - the building
blocks of quantum computing \cite {divincenzo1} - due to their discrete, atomic-like density of
states, the relative ease of generating single excitations (excitons or spins) per dot, and the
long lifetimes and coherence times of these excitations \cite {nirmal, kroutvar, flissikowski,
crookerAPL, gupta}. In epitaxially-grown quantum dots, prototype designs for generating,
manipulating, and detecting coherent superpositions of exciton eigenstates have been recently
reported \cite {bonadeo, chen, li}.  Of particular interest for quantum information processing are
entanglement schemes that exploit the distinct exciton eigenstates that occur naturally in
epitaxial dots due to shape anisotropy (elongation) of the dot and its associated anisotropic
exchange \cite {flissikowski, bonadeo}. This anisotropy mixes the lowest optically allowed (spin
$\pm1$) bright excitons, giving a ``fine structure" of two eigenstates, $|X\rangle = (|+1\rangle +
|-1\rangle)/\sqrt{2}$ and $|Y\rangle = (|+1\rangle - |-1\rangle)/\sqrt{2}$, linearly polarized
along the inequivalent $[110]$ and $[1\bar{1}0]$ substrate axes and typically split by 10-500
$\mu$eV \cite {gammon, nikitin, kulakovskii, puls, dzhioev, ivchenko}.

While experimental efforts have largely focused on epitaxially-grown quantum dot systems, colloidal growth \cite{murray} of nanocrystal quantum dots (NQDs) provides an alternative route toward quantum confinement in semiconductors. In contrast with epitaxial dots, NQDs are nearly spherical in shape and generally provide much stronger quantum confinement (radii down to 10 {\AA}), and are free from the straining influence of an adjacent substrate. Further, NQDs can exhibit near-perfect wurtzite crystallinity, so that confined electrons, holes, and excitons can have long lifetimes \cite{crookerAPL}, robust spin coherence \cite{gupta} and room temperature radiative quantum yields approaching unity \cite{peng}. Importantly, colloidal NQDs can also be chemically assembled, post-synthesis, into two- and three-dimensional
functional structures \cite{alivisatos2, redl}, with sizes and
shapes flexibly controlled during growth \cite{peng2}.

Despite the broad interest in exciton fine structure in epitaxially-grown dots, there are no reported studies of anisotropic exchange and
corresponding $|X,Y\rangle$ fine structure in colloidal NQDs.
Techniques which reveal these nondegenerate $|X,Y\rangle$ bright
eigenstates in {\it ensembles} of epitaxial dots (e.g.,
time-resolved quantum beats \cite {flissikowski} or optical
alignment-to-orientation \cite {dzhioev}), rely on the fact that
dots in an epitaxial ensemble share the same crystal axes and
direction of shape anisotropy. In typical colloidal ensembles,
however, the random orientation of the NQDs precludes similar
success. Alternatively, high-resolution, polarization-resolved
photoluminescence (PL) studies of {\it single} epitaxial dots,
which directly reveal the $|X\rangle$ and $|Y\rangle$ fine
structure splitting \cite{gammon, nikitin, kulakovskii}, are
hampered in studies of single colloidal NQDs by the effects of
spectral diffusion and blinking \cite{empedocles}. (Single-NQD
studies to date typically employ high spectral resolution {\it or}
polarization resolution, but not both. \cite {htoon, empedocles})

Here we demonstrate and measure $|X,Y\rangle$ bright exciton fine structure splittings in
randomly-oriented CdSe NQD ensembles through the use of ultrahigh magnetic fields {\bf B} to 33 T.
In these prolate NQDs, the lowest bright (spin $\pm1$) excitons possess dipoles that are nominally
degenerate in the plane normal to the wurtzite $\hat{c}$-axes of the NQDs \cite{empedocles} (the
spin $\pm2$ dark excitons are split off due to {\it isotropic}, or short-range, electron-hole
exchange \cite{nirmal, efros}). However, as in epitaxial dots, any deviation of the NQD shape away
from axial symmetry lifts this degeneracy, giving two bright eigenstates with dipoles directed
along the inequivalent $|X\rangle$ and $|Y\rangle$ semi-minor NQD axes \cite{ivchenko}. Consider
NQDs with $\hat{c} \parallel$ {\bf B} $\parallel \hat{z}$, as shown in Fig. 1a: When the magnetic
Zeeman energy ($g \mu_B B$) exceeds any fine structure splitting of bright excitons
($\Delta_{XY}$), the $|X\rangle$ and $|Y\rangle$ states evolve into eigenstates quantized along
{\bf B} \cite{kulakovskii, puls}, regardless of NQD orientation. In this limit, the states' spin
projections are quantized parallel and antiparallel to {\bf B} ($|\pm1\rangle_z$), and they couple
to circularly polarized light. The exciton eigenenergies do not evolve linearly with {\bf B} but
rather as $\pm \frac{1}{2} \sqrt{\Delta_{XY}^2 + (g\mu_B B)^2}$ (see Fig. 1a). In an ensemble of
NQDs, monitoring the energy difference between these levels provides a measure of the {\it
average} characteristic fine structure splitting of bright excitons.

\begin{figure}[tbp]
\includegraphics[width=.45\textwidth]{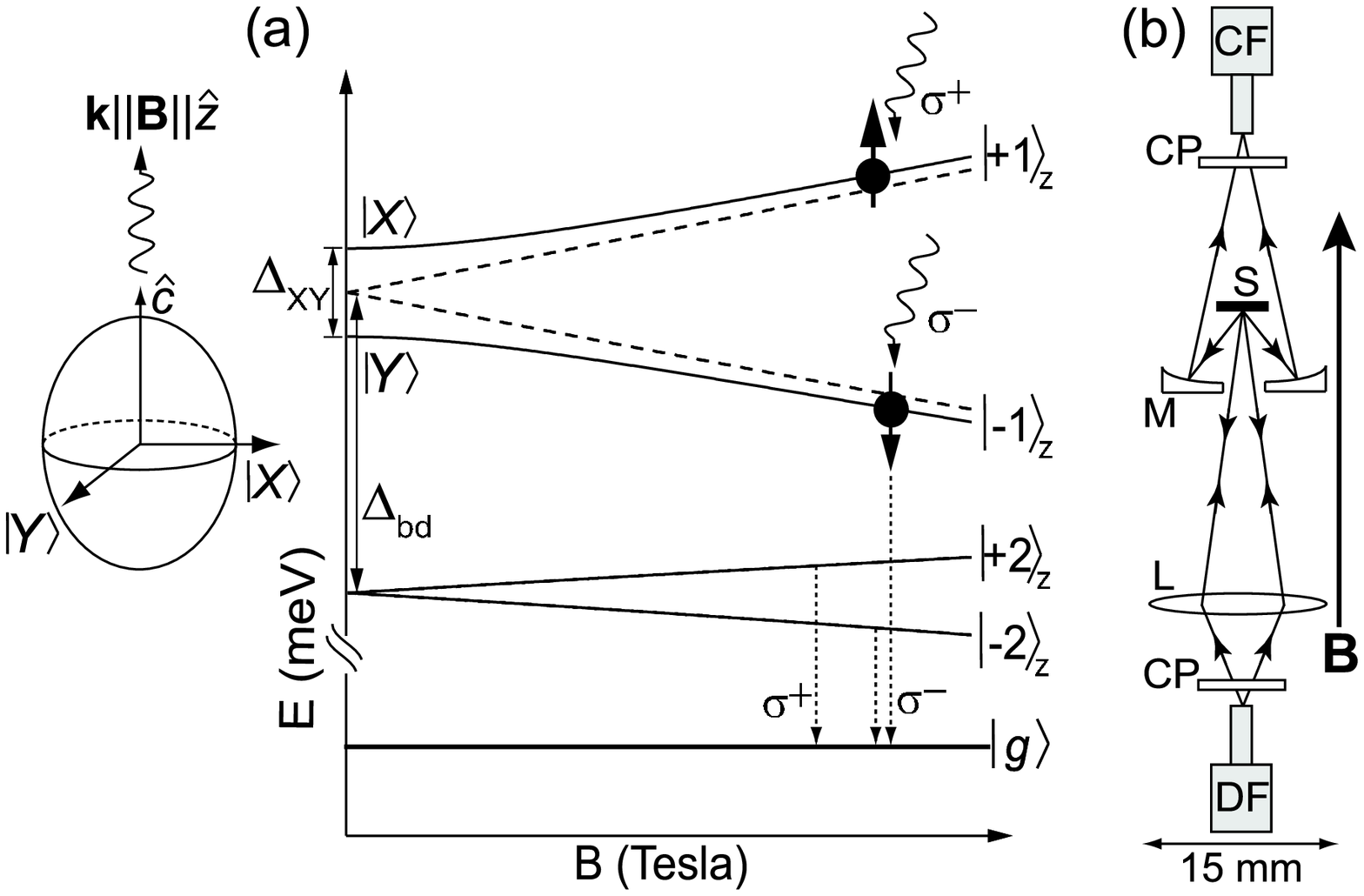}
\caption{(a) Spin-polarized resonant PL of CdSe NQDs in high magnetic fields {\bf B}. Spin-up or
-down bright excitons are resonantly excited with circularly polarized light, and the near-resonant
polarized PL from spin-up or spin-down excitons (bright or dark) is selectively analyzed. For NQDs
with $\hat{c} \parallel${\bf B}$\parallel \hat{z}$ as shown, all states are quantized along
$\hat{z}$ when the Zeeman energy exceeds any zero-field $|X,Y\rangle$ fine structure splitting
($g\mu_{B}B> \Delta_{XY}$). (b) The fiber-coupled probe: light from the delivery fiber (DF) is
focused by a lens (L) onto the sample (S). PL is focused onto the collection fiber (CF) by an
annular mirror (M). CP denotes circular polarizer.} \label{fig1}
\end{figure}

We use both steady-state and time-resolved spin polarized resonant PL spectroscopy (also known as
``fluorescence line narrowing" \cite{nirmal}) to selectively and resonantly excite spin-up or
spin-down bright excitons into the NQDs, and selectively measure the near-resonant emission from
spin-up or spin-down excitons. Above B$\sim$10 T, the spectra develop a narrow, circularly
polarized peak associated with spin-flipped bright excitons. The nontrivial evolution of this peak
with magnetic field directly reveals a large, average fine structure splitting of the bright
exciton doublet, of order 1-2 meV depending on NQD size. This $|X,Y\rangle$ splitting is confirmed
by polarization-resolved PL from {\it single} NQDs.

Samples of colloidal CdSe NQDs having $<$10\% size dispersion and mean radii ranging from 14 to 29
{\AA} were prepared by organometallic synthesis \cite {murray}. The NQDs were coated with a ZnS
shell for improved surface passivation and capped with trioctylphosphine oxide. Small volumes of
hexane/octane solution containing NQDs were drop-cast onto GaAs substrates, giving optical-quality
films of randomly oriented NQDs. Studies were performed at low temperatures to 1.7~K and in
magnetic fields up to 33 T at the National High Magnetic Field Laboratory. In the Faraday geometry
({\bf B} $\parallel$ {\bf k}), bright excitons were resonantly excited into the NQDs with a
narrowband tunable dye laser using a fiber-coupled probe which inserts directly into the bore of a
$^4$He cryostat \cite{furis}. The photon energy of this weak ($<$1 mW) laser is tuned to the
low-energy tail of the NQD absorption band, providing resonant excitation of the lowest bright
excitons in a sub-ensemble of NQDs of specific size \cite{nirmal} and preferred orientation
$\hat{c}\parallel$ {\bf B} \cite {orient}. We determine the specific NQD size directly from the
observed bright-dark exciton splitting $\Delta_{bd}$ \cite{nirmal}. The probe (Fig. 1b) uses
annular mirrors to avoid collection of scattered excitation light while allowing efficient
collection of near-resonant emission. This design permits clear resolution of features within 0.5
meV of the excitation laser energy using a 0.5 m single-axis spectrometer and a 2400 g/mm grating.
Thin-film circular polarizers allow selective excitation and detection of spin-up and spin-down
excitons. Time- and polarization-resolved PL spectra were obtained via time-correlated single
photon counting using 8 ns excitation pulses ``sliced" from the dye laser with an acousto-optic
modulator.

\begin{figure}[tbp]
\includegraphics[width=.43\textwidth]{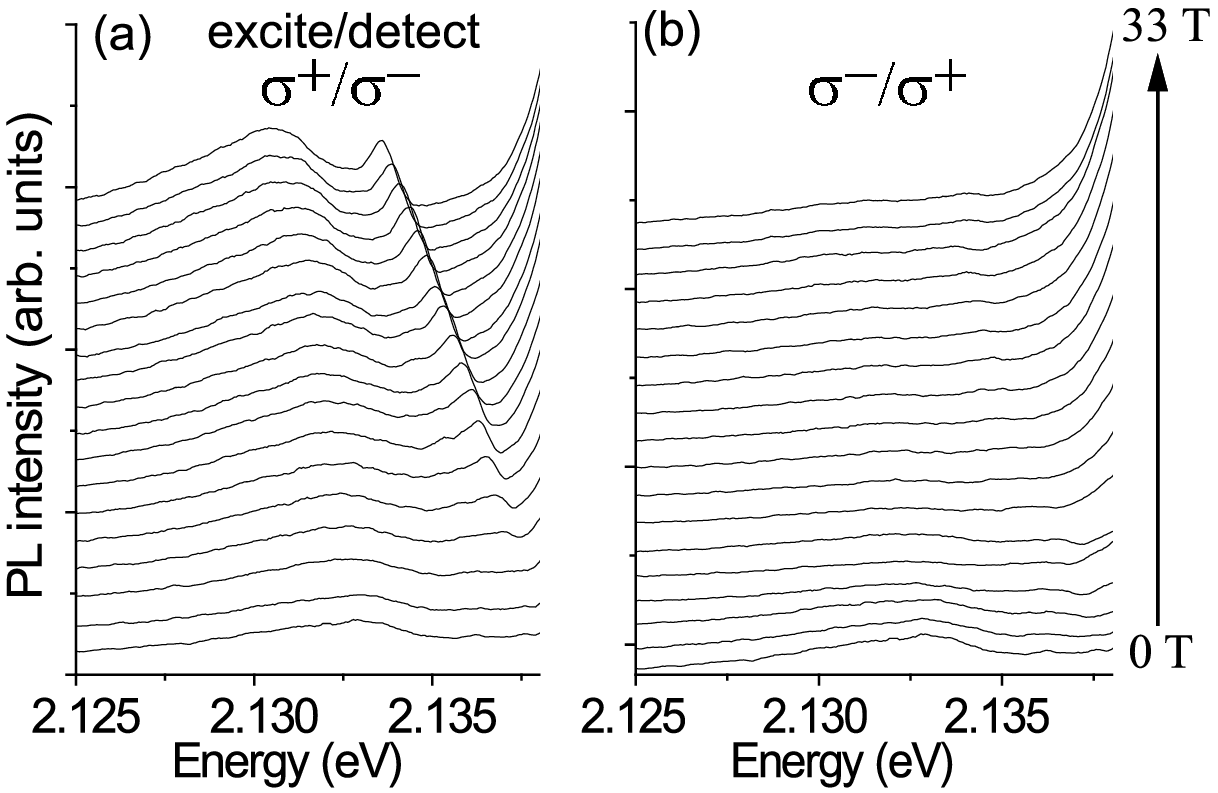}
\caption{Near-resonant, circularly polarized PL spectra from 19 {\AA} CdSe NQDs vs. magnetic field
at 1.7~K. (a) Exciting spin-up bright excitons at 2.1387 eV and detecting $\sigma^-$ emission from
spin-down excitons (both dark and bright).  A narrow peak develops above $\sim$10T. (b) Exciting
spin-down bright excitons and detecting $\sigma^+$ PL from spin-up excitons.} \label{fig2}
\end{figure}

Figure 2 shows the measured resonant emission from 19~{\AA} NQDs at 1.7~K, from 0 to 33 T for both
cross-circular polarizations (exciting/detecting $\sigma^+/\sigma^-$ and $\sigma^-/\sigma^+$).  At
B=0, emission from spin-2 (nominally ``dark") excitons is centered at 2.1330 eV, redshifted by
$\Delta_{bd}\sim$ 6 meV from the 2.1387 eV excitation laser ($\Delta_{bd}$ is the bright-dark
exciton splitting; see Fig. 1). With increasing {\bf B}, the dark exciton PL becomes strongly
$\sigma^{-}$ polarized, consistent with prior non-resonant magneto-PL studies of NQD ensembles
\cite {johnston-halperin}.

Most strikingly, the spectra develop a narrow, circularly polarized emission line above $\sim$10 T
when exciting $|+1\rangle_z$ excitons with $\sigma^+$ light and detecting $\sigma^-$ emission.
This narrow line appears between the excitation energy and the dark-exciton PL, and in agreement
with related studies on epitaxial quantum dot ensembles \cite {puls}, this line is absent in the
other three excitation/detection schemes ($\sigma^-/\sigma^+$, $\sigma^\pm/\sigma^\pm$). The
evolution of its energy shift from the excitation laser is not strictly linear and does not
extrapolate back to zero shift at B=0 (to be shown in Fig. 4). Significantly, emission from this
narrow peak is very short-lived (Fig. 3). At 15 T and 1.7 K, emission decays acquired on this peak
exhibit a fast initial component that tracks the 8 ns excitation pulse, whereas decays acquired
just off of the narrow peak exhibit only the slow dynamics characteristic of dark exciton PL.
These fast decays are consistent with the $\sim$10 ns radiative lifetime of bright excitons
\cite{crookerAPL} or a spin-flip resonant Raman process \cite {sapega}, and are not due to
inadvertent collection of the nearby excitation laser: at 20~K, both the new peak and the fast
initial decay disappear. Based on i) the emission energy of this narrow peak, ii) its
polarization, iii) its nontrivial evolution with applied magnetic field and iv) its comparatively
fast lifetime, we associate this feature with spin-down bright excitons ($|-1\rangle_z$). Thus,
the energy difference between the $\sigma^+$ excitation laser and the narrow $\sigma^-$ peak
reveals the average splitting between $|\pm1\rangle_z$ bright excitons in the NQD ensemble, from
which the intrinsic fine structure can be measured.

\begin{figure}[tbp]
\includegraphics[width=.43\textwidth]{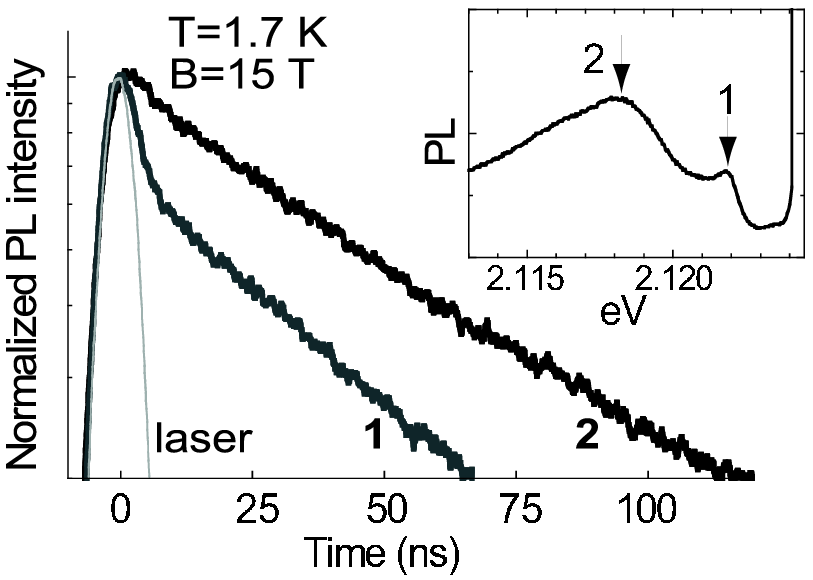}
\caption{Time-resolved PL at 1.7~K and 15 T, using $\sigma^+$ pulsed resonant excitation at 2.125
eV and $\sigma^-$ detection on the narrow emission line (1) and on the dark exciton PL (2).}
\label{fig3}
\end{figure}

Figure 4(a) shows the measured energy splitting between $|\pm1\rangle_z$ bright excitons versus
{\bf B} for four distinct NQD sizes. A linear extrapolation of these high-field data back to B=0
does not, however, intercept the origin. Moreover, the slope of the data flattens at low fields
(5-15 T; seen most clearly in the 18 {\AA} NQDs). As discussed in Fig. 1a, these trends directly
indicate the presence of anisotropic exchange and an intrinsic fine structure splitting of bright
excitons in NQDs. The narrow emission peak only develops above B$\sim$10 T, when $g\mu_B
B>\Delta_{XY}$ and the $|\pm1\rangle_z$ states are well defined. At low fields ($g\mu_B
B<\Delta_{XY}$), spin projection along $\hat{z}$ is not a good quantum number, and circularly
polarized light does not selectively excite spin-oriented excitons.

In a simplified Hamiltonian containing the lowest four dark and
bright excitons ($|\pm2\rangle$, $|\pm1\rangle$), anisotropic
exchange leads to off-diagonal terms that mix the $|\pm 1 \rangle$
states at B=0 \cite {ivchenko}, giving linear $|X\rangle$ and
$|Y\rangle$ eigenstates split by $\Delta_{XY}$ (see Fig. 1a). In
this framework the eigenenergies evolve as $\pm \frac{1}{2}
\sqrt{\Delta_{XY}^2 + (g\mu_B B)^2}$. Averaging this dependence
over all NQD orientations \cite{orient}, the lines in Fig. 4(a)
show the predicted evolution of the bright exciton splitting for
18 and 24 {\AA} NQDs.  The data are fit very well using only two
free parameters: $\Delta_{XY}$ and the hole g-factor. Electron
g-factors are obtained from spin precession studies \cite {gupta}.
As shown in Fig. 4b, the fits reveal that $\Delta_{XY}$ increases
systematically from 1.1 to 2.0 meV as the NQDs shrink in size from
R=29 to 14 {\AA}. $\Delta_{XY}$ scales (very approximately) with
the inverse NQD volume, in rough accord with theories that ascribe
$\Delta_{XY}$ to the long-range, anisotropic electron-hole
exchange interaction \cite {ivchenko}. By analogy with epitaxial
quantum dot systems, the origin of $\Delta_{XY}$ in these wurtzite
NQDs likely arises from a structural asymmetry of the confining
potential resulting in inequivalent $|X\rangle$ and $|Y\rangle$
semi-minor axes of the NQD (\emph{i.e.}, a deviation from
$\hat{c}$-axis cylindrical symmetry). In this case, the bright
exciton Hamiltonian contains additional, off-diagonal terms of the
form $J_{X}^{2}-J_{Y}^{2}$ \cite {goupalov3}, where $J_{X,Y}$
represent hole spin projections along the inequivalent semi-minor
NQD axes. The average splitting is approximately
an order of magnitude larger than values measured in CdSe
epitaxial quantum dots \cite{nikitin,kulakovskii,puls}, in accord with the expected magnitude of the Coulomb matrix elements (arising from electron-hole exchange), which are inversely proportional to quantum dot volume \cite{ivchenko}. In comparison with typical disk-shaped epitaxial quantum dots (100 {\AA} diameter, 20 {\AA} height), spherical NQDs having 18-24 {\AA} radii are $\sim$3-7 times smaller in volume.

\begin{figure}[tbp]
\includegraphics[width=.45\textwidth]{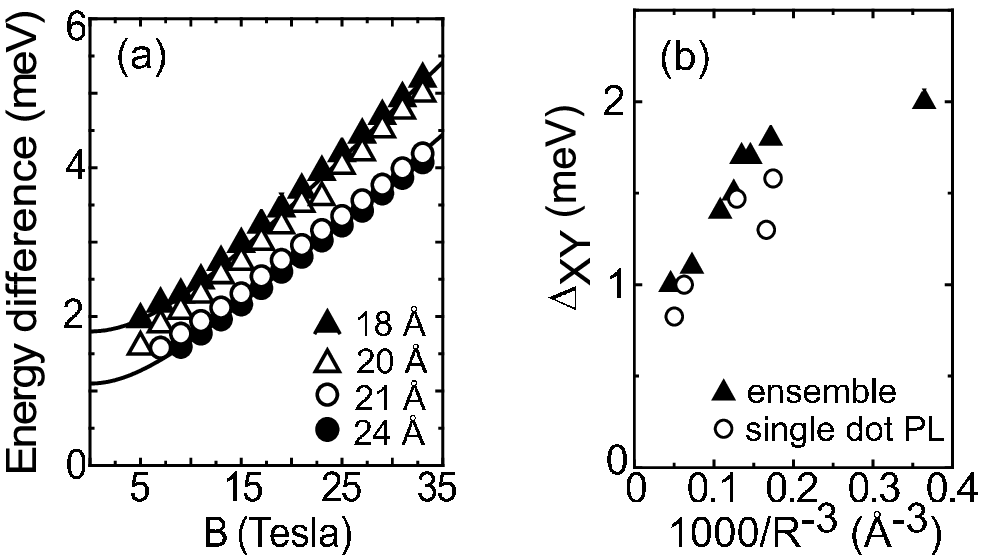}
\caption{(a) The energy difference between the pump laser and the narrow emission peak vs. {\bf B}
for four NQD sizes. Lines are fits using an anisotropy model (see text). (b) The measured
zero-field bright exciton fine structure splitting, $\Delta_{XY}$, vs. inverse NQD volume (the
fits give $\pm$20\% error.) Different NQD sizes are probed (within a given ensemble) by tuning the
pump laser.} \label{fig4}
\end{figure}

Finally, we independently confirm the $|X,Y\rangle$ fine structure and anisotropic exchange in
NQDs through high-resolution, polarization-resolved, low-temperature nonresonant PL of single
NQDs. To minimize the effects of spectral diffusion and blinking, we \emph{simultaneously} detect
both orthogonal, linearly polarized PL components (labeled $x'$ and $y'$) by using a polarizing
beamsplitter in front of an imaging spectrometer. A half-wave plate allows effective rotation of
the ($x',y'$) detection axes. At 4K, $\sim$10\% of the surveyed NQDs exhibit narrow PL lines and a
resolvable splitting between orthogonal linearly-polarized PL components. Figure 5(a) shows direct
spectroscopic evidence for this $|X,Y\rangle$ fine structure of bright excitons in one such NQD.
With the $x'$ and $y'$ detection axes aligned with this NQD's particular $|X\rangle$ and
$|Y\rangle$ emission axes, two clear PL peaks are observed, orthogonally polarized and split by
$\Delta_{XY} \cong 0.8$ meV. For a NQD of this size (R=26.5{\AA}), this value is in reasonable
agreement with $\Delta_{XY}$ inferred from the high-field resonant PL of NQD ensembles (open
symbols in Fig. 4b show $\Delta_{XY}$ measured from single NQDs). Upon rotating the ($x',y'$)
detection axes by 40$^\circ$, both polarizations now show a double-peaked feature, as expected
[Fig. 5(b)]. A further 40$^\circ$ rotation brings the detection axes (nearly) again into alignment
with the NQD's intrinsic $|X\rangle$ and $|Y\rangle$ emission axes [Fig. 5(c)], but with reversed
polarizations. The spectra are not artifacts of multiple NQD emission: In the presence of a
significant redshift due to spectral diffusion [Fig. 5(b)], the spectral shape and relative
splitting of the two polarized PL peaks are unchanged.

\begin{figure}[tbp]
\includegraphics[width=.45\textwidth]{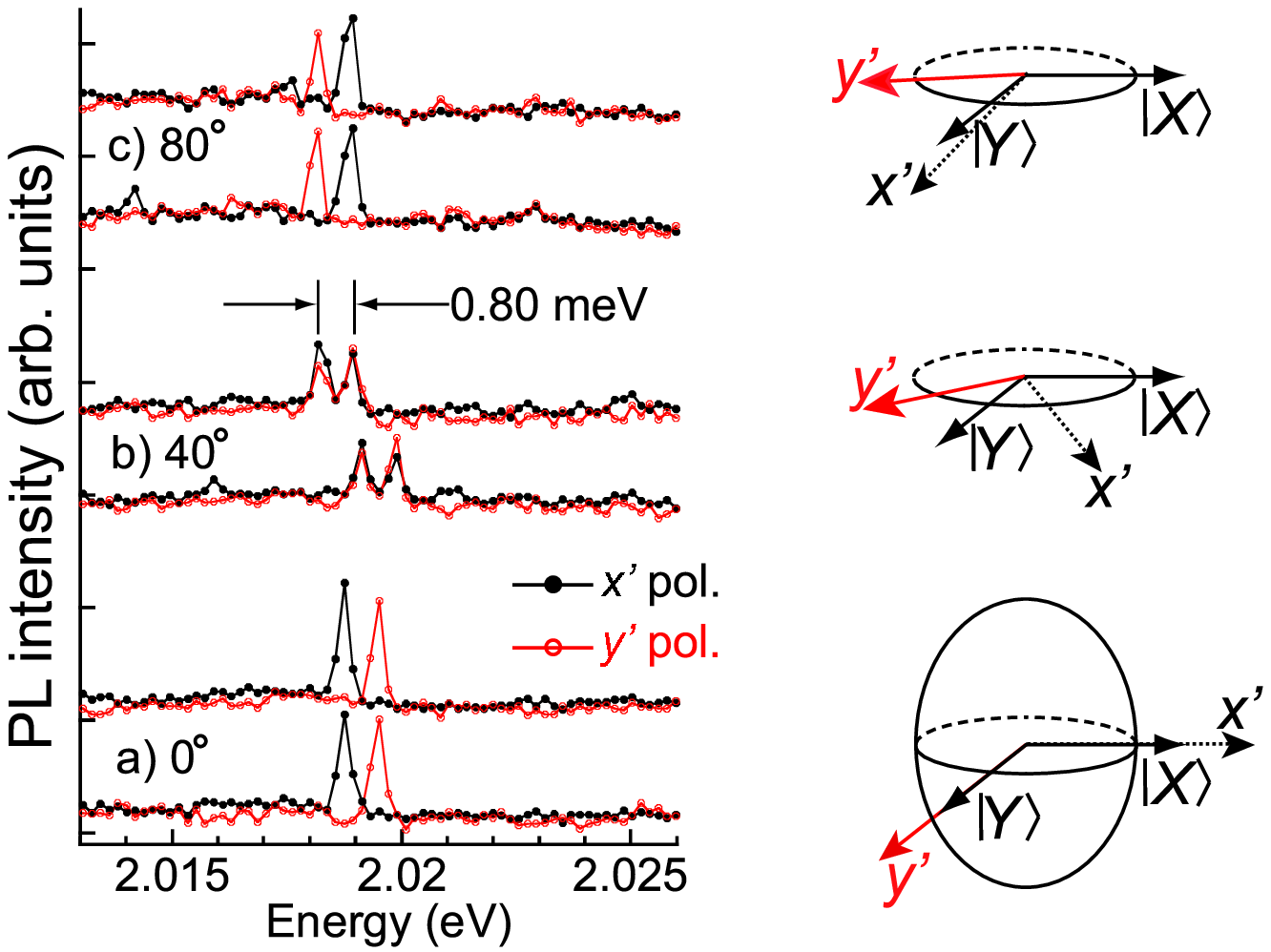}
\caption{(color online) Polarization-resolved bright exciton PL from a single CdSe NQD at
T$\sim$4K. Two 20 s exposures are shown for each orientation of the linear detection axes $x'$ and
$y'$: a) 0$^\circ$, b) 40$^\circ$, and c) 80$^\circ$ with respect to the NQD's intrinsic
$|X\rangle$ and $|Y\rangle$ emission axes.  Spectral diffusion is clearly seen in b).}
\label{fig5}
\end{figure}

In conclusion, ultrahigh magnetic fields to 33 T and
spin-polarized resonant PL reveal the intrinsic, average
zero-field $|X,Y\rangle$ fine structure splitting of bright
excitons in CdSe NQD ensembles. We find $\Delta_{XY}$ ranges from
1.1-2.0 meV (depending on NQD size), considerably larger than
corresponding average splittings typically measured in epitaxial
quantum dots. Low-temperature PL from single colloidal NQDs,
employing {\it both} polarization and high spectral resolution,
also directly reveals the intrinsic $|X,Y\rangle$ fine structure
splitting of bright excitons. In analogy with epitaxial quantum
dot systems, this additional exciton fine structure may prove
beneficial for future quantum information processing schemes that
exploit the flexibility and functionality of chemically assembled
colloidal nanocrystals.

We gratefully acknowledge S. Goupalov, A. Efros, and P. Robbins for valuable discussions. This
work was supported by the Los Alamos LDRD program and the DOE Office of Basic Energy Sciences.


\newpage

\end{document}